\RequirePackage{ifpdf}
\ifpdf 
\documentclass[pdftex]{sigma}
\else
\documentclass{sigma}
\fi

\begin{document}
\allowdisplaybreaks

\renewcommand{\PaperNumber}{002}

\FirstPageHeading

\ShortArticleName{On the Effective Action of Dressed Mean Fields
for  ${\cal N} =4$ Super-Yang--Mills Theory}

\ArticleName{On the Effective Action of Dressed Mean Fields\\ for
$\boldsymbol{{\cal N} =4}$ Super-Yang--Mills Theory}

\Author{Gorazd CVETIC~$^\dag$, Igor KONDRASHUK~$^{\dag\ddag}$ and
Ivan SCHMIDT~$^\dag$} \AuthorNameForHeading{G. Cvetic, I.
Kondrashuk and I. Schmidt}

\Address{$^\dag$~Departamento de F\'\i sica, Universidad T\'ecnica Federico Santa Mar\'\i a, \\
$\phantom{^\dag}$~Avenida Espa\~{n}a 1680, Casilla 110-V,
Valparaiso, Chile}
\EmailD{\href{mailto:gorazd.cvetic@usm.cl}{gorazd.cvetic@usm.cl},
\href{mailto:igor.kondrashuk@usm.cl}{igor.kondrashuk@usm.cl},
\href{mailto:ivan.schmidt@fis.utfsm.cl}{ivan.schmidt@fis.utfsm.cl}}
\Address{$^\ddag$~Departamento de Ciencias Basicas, Universidad del Bio-Bio,\\
$\phantom{^\ddag}$~Campus Fernando May, Casilla 447, Avenida
Andreas Bello, Chillan, Chile}

\ArticleDates{Received October 30, 2005, in final form January 01,
2006; Published online January 09, 2006}

\Abstract{On the basis of the general considerations such as
$R$-operation and Slavnov--Taylor identity we show that the
effective action, being understood as Legendre transform of the
logarithm of the path integral, possesses particular structure in
${\cal N} =4$ supersymmetric Yang--Mills theory for kernels of the
effective action expressed in terms of the dressed effective
fields. These dressed effective fields have been introduced in our
previous papers as actual variables of the effective action. The
concept  of dressed effective fields naturally appears in the
framework of solution to Slavnov--Taylor identity. The
particularity of the structure is independence of these kernels on
the ultraviolet regularization scale $\Lambda.$ These kernels are
functions of mutual spacetime distances and of the gauge coupling.
The fact that  $\beta$ function in this theory vanishes is used
significantly.}


\Classification{81Q30}

\def\a{\alpha}
\def\b{\beta}
\def\db{{\dot{\beta}}}
\def\t{\theta}
\def\bt{{\bar{\theta}}}
\def\yb{{\bar{y}}}
\def\Db{{\bar{D}}}
\def\Tr{{\rm Tr}}
\def\dis{\displaystyle}
\def\le{\left(}
\def\ri{\right)}
\def\da{{\dot{\alpha}}}
\def\no{\nonumber}
\def\del{\delta}
\def\bW{\bar{W}}
\def\G{\Gamma}
\def\rar{\rightarrow}
\def\fra1g2{\frac{1}{g^2}}
\def\dg{{\dagger}}
\def\Vt{\tilde{V}}
\def\Kt{\tilde{K}}
\def\SQa{\sqrt{\tilde{\a}}}
\def\e{\epsilon}
\def\sm{{\sigma_m}}
\def\f12{\frac{1}{2}}
\def\F{\Phi}
\def\pd{\partial}
\def\ve{\varepsilon}
\def\F{\Phi}
\def\bF{\bar{\Phi}}

Slavnov--Taylor (ST) identity \cite{ST} is an important tool in
quantum field theory. It is a consequence of BRST symmetry
\cite{BRST} of the tree level action of gauge theories, and it
consists in an equation written for a functional that is called
effective action \cite{SF}. An approach to solving the ST identity
in gauge theories has been proposed recently
\cite{Cvetic:2002dx,Cvetic:2002in}. In this Letter we re-consider
our analysis for a particular case of   ${\cal N} =4$
supersymmetric theory. Our analysis will be based on five
theoretical tools: $R$-operation \cite{BoSh}, gauge symmetry,
${\cal N} =4$ supersymmetry, the ST identity itself, and absorbing
two point Green's functions into a re-definition of the effective
fields. Effective fields are variables of the effective action
\cite{SF}.

${\cal N} =4$ super-Yang--Mills theory is widely considered  from
the point of view AdS/CFT correspondence \cite{Maldacena:1997re}.
Anomalous dimensions of gauge invariant operators are related to
energies of string states \cite{Berenstein:2002jq}. In this Letter
we consider ${\cal N} =4$ super-Yang--Mills theory from a
different point. We analyze one particle irreducible (proper)
correlators of this theory, which are kernels of the effective
action. For example, a kernel can be proper vertex of several
gluons. We hope this analysis can have application to calculation
of maximal helicity violating amplitudes of processes with $n$
gluons \cite{Witten:2003nn,Bern:1994zx}. ${\cal N} =4$
super-Yang--Mills theory is useful theoretical playground to
understand better the problems that stand in QCD. This model has
special particle contents. In addition to one gluon and four
Majorana fermions it contains six scalar fields. All the particles
are in the adjoint representation of $SU(N)$ gauge group.

It has been  known for a long time that for ${\cal N} =4$
Yang--Mills supersymmetric theory beta function of gauge coupling
vanishes \cite{Ferrara:1974pu,Jones:1977zr,
Avdeev:1980bh,Sohnius:1981sn}. We extensively use this fact in our
analysis. The basic notation of this paper coincides with notation
of~\cite{Cvetic:2002dx}. The main ST identity is \cite{SF}
\begin{gather}
 \Tr\left[\int dx\, \frac{\del \G}{\del A_m(x)}\frac{\del \G}{\del K_m(x)}
  + \int dx\, \frac{\del \G}{\del c(x)}\frac{\del \G}{\del L(x)}
  - \int dx\, \frac{\del \G}{\del b(x)}
   \le\frac{1}{\a} \pd_m A_m(x)\ri \right] \nonumber \\
\qquad{} + \int dx\, \frac{\del \G}{\del \phi(x)} \frac{\del
\G}{\del k(x)} + \int dx\, \frac{\del \G}{\del \bar{k}(x)}
\frac{\del \G}{\del\bar{\phi}(x)} = 0. \label{main}
\end{gather}
The effective action is a functional of all the effective fields
and external sources participating in this equation,  $\G \equiv
\G[A_m,b,c,\phi,\bar{\phi},K_m,L,k,\bar{k}]$ \cite{Lee,SF}. The
external sources $K_m$, $L$, $k$, $\bar{k}$ are coupled in the exponential
of the path integral to the BRST transformations of fields from
the measure of the path integral \cite{SF}, that is, to the BRST
transformations of fields $A_m$, $c$, $\phi$, $\bar{\phi}$ of the tree
level action, respectively. The effective fields  $A_m$, $b$, $c$,
$\phi$, $\bar{\phi}$ are traditionally designated by the same
letters as are  the fields  $A_m$, $b$, $c$, $\phi$, $\bar{\phi}$,
which are variables of the path integral. The effective fields are
defined as variational derivatives of the logarithm of the path
integral with respect to the corresponding external sources
coupled to these
 variables
of integration in the path integral~\cite{SF}.
 The matter effective
field $\phi$ stands for spinors as well as for scalars. We assume
summations over all indices of the representation of matter
fields. The traditional Lorentz gauge fixing is taken and the
corresponding Faddeev--Popov ghost action introduced according to
the line of Ref.~\cite{Avdeev:1980bh}.  These terms break
supersymmetry of the tree level action. The $\b$ function is zero
but anomalous dimensions of propagators are non-zero
\cite{Avdeev:1980bh}.

Consider the vertex $Lcc$. Here we do not specify arguments of the
effective fields. It is the only vertex, which is invariant with
respect to the ST identity at the classical level. At the quantum
level it transforms to the form
\begin{gather}
\langle Lcc\rangle  \times \langle Lcc\rangle  + \langle
LccA\rangle \times \langle K_m\pd_mc\rangle  = 0.  \label{rough}
\end{gather}
This is a direct consequence of the main ST identity (\ref{main})
and is a schematic form of the ST identity relating the $Lcc$ and
$LccA$ field monomials. The precise form of this relation can be
obtained by differentiating the identity~(\ref{main}) with respect
to $L$ and three times with respect to $c$ and then by setting all
the variables of the effective action equal to zero. The brackets
in~(\ref{rough}) mean that we have taken functional derivatives
with respect to fields in the corresponding brackets at different
arguments and then have put all the effective fields equal to
zero.

We know from the theory of $R$-operation \cite{BoSh} that in
Yang--Mills theory the divergences  can be removed by re-defining
the fields and the gauge coupling. Thus, there are four
renormalization constants that multiply  the ghost, gluon, spinor,
and scalar fields \cite{SF}. The gauge coupling also must be
renormalized but this is not the case in the theory under
consideration.  In this paper we concentrate on two
regularizations: regularization by higher derivatives described
in~\cite{SF}, $\Lambda$ is the regularization scale, and
regularization by dimensional reduction~\cite{Siegel:1979wq}. The
regularization by higher derivatives  has been constructed for
supersymmetric theories
in~\cite{Krivoshchekov:1978xg,West:1985jx}. Having used this
regularization, new scheme has been proposed in
\cite{Slavnov:2003cx,Slavnov:2002kg,Slavnov:2001pu}. We assume
here that the component analog of that scheme can be constructed.
The regularization by higher derivatives  provides strong
suppression of ultraviolet divergences by introducing  additional
terms with higher degrees of covariant derivatives acting on
Yang--Mills tensor into the classical action, which are suppressed
by appropriate degrees of the regularization scale $\Lambda.$  In
addition to this, it is necessary to introduce a modification of
the Pauli--Villars regularization to guarantee the convergence of
the one-loop diagrams~\cite{SF}. This scheme does not break gauge
invariance beyond one loop level. Moreover, it has been suggested
in~\cite{SF} that such modification by Pauli--Villars terms to
remove one-loop infinities is gauge invariant by construction. To
regularize the fermion cycles, the usual Pauli--Villars
regularization can be used. However,  this approach, when applied
to explicit examples, is known to yield incorrect results in
Landau gauge \cite{Martin:1994cg}. A number of suggestions have
been put forward to treat this problem
\cite{Bakeyev:1996is,Asorey:1995tq}. As was shown
in~\cite{Asorey:1995tq}, the contradiction, noticed in
\cite{Martin:1994cg} is related to the singular character of
Landau gauge. In all other covariant gauges the method works, and
to include also the Landau gauge one has to add one more
Pauli--Villars field to get the correct result
\cite{Bakeyev:1996is,Asorey:1995tq}. We analyze the theory in the
regularization by the dimensional reduction in a parallel
way~\cite{Siegel:1979wq}.

At one-loop level the part associated with the divergence of $Lcc$
term must be invariant itself under the ST identity since the
second term in identity (\ref{rough}) is finite in the limit of
removing regularization $\Lambda \rar \infty$
\cite{Cvetic:2002dx}. According to Ref. \cite{Cvetic:2002dx}, this
results in the following integral equation for the part of the
correlator $Lcc$ corresponding  to the superficial divergence
$\sim \ln\frac{p^2}{\Lambda^2}$:
\begin{gather}
\int dx \, \G_{\Lambda}(y',x,z')\G_{\Lambda} (x,y,z) = \int dx\,
\G_{\Lambda}
(y',y,x)\G_{\Lambda} (x,z,z') \nonumber\\
\qquad{} = \int dx\, \G_{\Lambda} (y',x,z)\G _{\Lambda} (x,z',y),
\label{integral}
\end{gather}
where $\G_{\Lambda} (x,y,z)$ is this scale($\Lambda-$) dependent
part of the most general parametrization $\G (x,y,z)$ of the
correlator $Lcc$,
\begin{gather}
\G \sim \int dx\,dy\,dz\, \G
(x,y,z)f^{abc}L^{a}(x)c^{b}(y)c^{c}(z). \label{A1}
\end{gather}
Here $f^{abc}$ is the group structure constant. The only solution
to the integral equation (\ref{integral}) is \cite{Cvetic:2002dx}
\begin{gather}
\G_{\Lambda}(x,y,z) = \int dx'\, G_{c}(x'-x)~G^{-1}_{c}(x'-y)
     G^{-1}_{c}(x'-z).        \label{result}
\end{gather}
The subscript $\Lambda$ means scale-dependent part of the
correlator. As can be seen, all scale-dependence of this
correlator is concentrated in the dressing function. The complete
correlator $Lcc$ at one loop level can be then written as
\begin{gather*}
\int dx\,dy\,dz\, \G (x,y,z)\frac{i}{2}f^{bca}
   L^{a}(x)c^{b}(y)c^{c}(z)  \\
\qquad {} = \int dx'\, dy'\, dz'\, dx\, dy\, dz\, \tilde{\G}
(x',y',z')G_{c}(x'-x)G^{-1}_{c}(y'-y)
\\
\qquad \phantom{=}{} \times G^{-1}_{c}(z'-z)\frac{i}{2}f^{bca}
L^{a}(x)c^{b}(y)c^{c}(z).
\end{gather*}
Here $\tilde{\G} (x',y',z')$ is scale-independent kernel of $Lcc$
correlator\footnote{In \cite{Cvetic:2002dx} we conjectured, on the
basis of equations~(\ref{integral}) and (\ref{A1}), that the
complete $\Gamma(x,y,z)$ has the structure (\ref{result}); this
would correspond to $\tilde \Gamma(x',y',z') \propto \delta(x' -
y') \delta(x' - z')$.}.

We absorb this dressing function $G_{c}$ into the corresponding
re-definition of the fields $L$ and $c$, and then divide the ghost
propagator into two parts, one of which is related to the dressing
function of the ghost field $G_c$, and we call another the
dressing function of the gluon field $G_A$. The effective field
$K$ and the antighost field $b$ get opposite re-definition by
integrating with the dressing function
$G^{-1}_{A}$~\cite{Cvetic:2002dx}. The important point here is
covariance of the part of the ST identity without gauge fixing
term with respect to such redefinitions \cite{Cvetic:2002dx}. In
terms of the dressed effective fields, we have a useful relation
which is a consequence of the main ST identity (\ref{main}) and
can be obtained by differentiating  the main ST identity two times
with respect to $\tilde{c}$ and one time with respect to
$\tilde{b}$. This resulting identity is
\begin{gather}
\langle \tilde{A}_m \tilde{b}\tilde{c}\rangle  \times \langle
\tilde{K}_m \tilde{c}\rangle  + \langle
\tilde{L}\tilde{c}\tilde{c}\rangle \times \langle \tilde{b}
\tilde{c}\rangle  = 0.  \label{rough2}
\end{gather}
Again, this identity is written in a schematic way. However, a new
important point appears here. Namely, since the two-point proper
functions  in terms of dressed effective fields are trivial tree
level two-point proper functions, the divergences of $\langle
\tilde{A}_m \tilde{b}\tilde{c}\rangle $ and $ \langle
\tilde{L}\tilde{c}\tilde{c}\rangle $ coincide. Since $ \langle
\tilde{L}\tilde{c}\tilde{c}\rangle $ is scale-(or $\Lambda$-)
independent, the $\langle \tilde{A}_m \tilde{b}\tilde{c}\rangle $
is scale-independent also.  Concerning the gluon propagator,  one
part of the divergence is in the dressing function $G_A,$ and the
rest of divergence would be absorbed in the re-definition of the
gauge coupling constant. This last divergence is absent in ${\cal
N} =4$ theories. All the other correlators are solved by the ST
identity in terms of the dressed effective fields and their
kernels are finite (do not possess divergence in the limit of
removing regularization) and scale-independent.

Infinite parts of the dressing functions will be one loop
counterterms corresponding to the re-definition of the fields.
Then we can repeat this procedure at two loop level and so on, up
to any order in loop number.  Indeed,  re-definition by
multiplication of the fields of the tree level action results in
re-defining external legs of proper correlators in comparison with
the unrenormalized theory. This property was used by Bogoliubov
and Shirkov \cite{BoSh} in the derivation of renormalization group
equations. Re-defining the effective fields by dressing functions
does not bring new aspects in this sense. Indeed, re-defining
variables of the path integral by dressing will result in the
dressing of the external legs of the proper correlators. By proper
correlators we mean kernels of $\G,$ that is, these kernels are
one particle irreducible diagrams. We have already used this
re-definition in the dressing functions in the
$\tilde{L}\tilde{c}\tilde{c}$ correlator. Thus, new superficial
divergences will have to satisfy the integral equation
(\ref{integral}) at two loop order too in terms of effective
fields  $\tilde{L}\tilde{c}\tilde{c}$, which are effective fields
dressed by one loop dressing function~$G_c.$ We then repeat this
re-definition in the same procedure in each order of the
perturbation theory as we did in the previous paragraph at one
loop level.

Until now the pure gauge sector has been considered. Fermions are
necessary for providing supersymmetry. Consider vertex $kc\phi$ at
one loop level. The superficial divergence of this vertex is
cancelled by the divergence of the vertex $Lcc$. This means that
in the part of the correlator  $kc\phi$ corresponding to
superficial divergence  $\sim \ln\frac{p^2}{\Lambda^2}$ this
divergence can be absorbed into the dressing functions in the
following way:
\begin{gather}
\int dx\,dx_1\,dy_1\,dy_2\, G_{\phi}(x-x_1)~G^{-1}_{c}(x-y_1)
     G^{-1}_{\phi}(x-y_2) k(x_1) c(y_1) \phi(y_2).
   \label{ex4}
\end{gather}
Note that the dressing functions for the fermions and scalars
$G^{-1}_{\phi}(x)$ are not fixed yet. We set them equal to halves
of the two point matter functions. The remaining vertices are
restored in the unique way because the ST identity works.

This theory has intrinsic on-shell infrared divergences, like
those canceled by brems\-stra\-hlung of soft gluons (such a
cancellation happens on shell). To regularize these divergences we
can introduce mass parameter $\mu$ \cite{Peskin}. Such a trick
breaks the ST identity by terms dependent on~$\mu.$ At the end the
dependence on~$\mu$ will disappear in physical matrix elements. We
mean by the physical matrix element a connected diagram on-shell
contribution to amplitudes of particles.  In comparision, the
dimensional reduction regulates both the ultraviolet and the
infrared dive\-gen\-ces. On the infrared side, this is better than
to introduce small mass parameter $\mu$, the reason being that in
a non-Abelian gauge theory the limit for infrared regulator going
to zero is in general singular. Infrared divergences cancel in the
physical matrix elements in dimensional reduction too. However,
one can think that in the effective action (off shell) they could
be  present since cancellation of the infrared divergences happens
(on shell) between proper and one particle reducible graphs
\cite{Peskin}. Below we will indicate that off shell these
infrared divergences do not exist at all in the position
space\footnote{We treat the theory in the position space. Infrared
divergences are absent off shell also in the momentum space by the
same reasons.}.

In principle, infrared divergences in the effective action
represent an outstanding problem that is not treated in the
present work in necessary details. We show in this paragraph that
in the regularization by the dimensional reduction this problem
does not appear.  It is enough to show this for the $Lcc$
correlator, because other correlators can be expressed in terms of
that one by ST identity, if we work in terms of dressed effective
fields. The one-loop contribution shown in Fig.~1 (the only
diagram that can be drawn) is apparently convergent in the Landau
gauge. The point is that the derivatives can be integrated out of
the graph due to the property of transversality of the gluon
propagator in the Landau gauge that immediately makes  this graph
convergent in the ultraviolet region, but it is safe in the
infrared. Note that in other gauges this correlator remains
divergent in the ultraviolet, and its scale dependence is
contained in the dressing function $G_c.$
\begin{figure}[t]
\centerline{\includegraphics[width=5.cm]{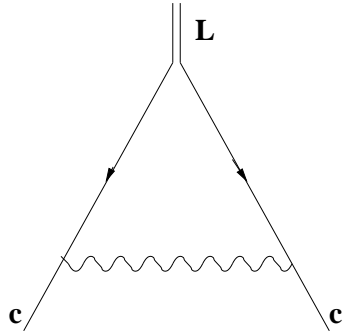}}
\caption{One-loop contribution to the $Lcc$ vertex. The wavy lines
represent the gluons, the straight line are for the ghosts.}
\end{figure}
The two-loop diagrams (planar) are drawn below in Fig.~2.
\begin{figure}[t]
\centerline{\includegraphics[width=5.5cm]{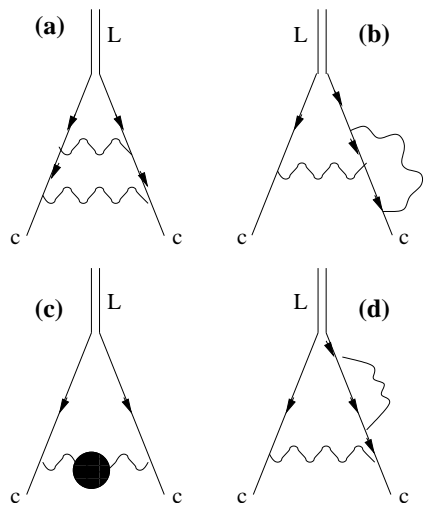}}
\caption{Two-loop diagrams for the $Lcc$ vertex. The wavy lines
represent the gluons, the straight lines the ghosts. The black
disc in (c) is for one-loop contribution in the renormalization of
the vector propagator from scalar, spinor and ghost fields.}
\end{figure}
The first two diagrams are apparently convergent in ultraviolet in
Landau gauge since all subgraphs are convergent. This is due to
the property of transversality that again allows to integrate out
the derivatives. The scale dependence in the diagram $(c)$ cancels
the scale dependence of diagram~$(d)$ because of ${\cal N} =4$
supersymmetry (absence of the gauge coupling renormalization).
Infrared region is also not dangerous since even in Landau gauge
the gluon propagator is safe in the infrared region.

However, no invariant regularization scheme is known for
supersymmetric theories, due to the well-known $\gamma_5$-problem:
on general grounds finite renormalizations have to be recursively
performed order by order in the loop expansion in order to
preserve the relevant functional identities \cite{Siegel:1980qs}.
In this paragraph we discuss how one can treat this problem by
using the technique described in this paper. The point is that the
vertex $Lcc$ is always convergent superficially in the Landau
gauge. In ${\cal N} = 4$ super-Yang--Mills theory ultraviolet
divergences in the subgraphs of the $Lcc$ vertex should cancel
each other at the end. The insertion of the operator of the
conformal anomaly into vacuum expectation values  of operators of
gluonic fields at different points in spacetime is proportional to
the beta function of the gauge coupling \cite{Collins:1976yq}. Due
to the algebra of the four-dimensional supersymmetry the beta
function should be zero \cite{Sohnius:1981sn}. Algebra of the
supersymmetry operators in the Hilbert space created by dressed
fields can be considered as four-dimensional in Lorentz indices as
well as in spinor indices since the limit $\e \rar 0$ is
non-singular at one-loop order, two-loop order and higher orders
as we have seen in the previous paragraphs. Thus, we can consider
each correlator as pure four-dimensional, solving  order-by-order
the problem of dimensional discrepancy of convolutions in Lorentz
and spinor indices.

The behavior of the theory in the IR region is not spoiled by the
higher derivative regularization too. This is clear from the
structure of the gluon propagator (in Landau gauge, for example)
\cite{SF}:
\begin{gather*}
D^{ab}_{\mu\nu} = \del^{ab}\left[-\le g_{\mu\nu} - \frac{k_\mu
k_\nu}{k^2} \ri\frac{1}{k^2 + k^6/\Lambda^4}\right].
\end{gather*}
Sixth degree of momentum in the denominator improves significantly
ultraviolet behavior but in the infrared it is negligible in
comparison with second degree of momentum.  The infrared
divergence appears when we en force put  the on-shell condition
$p_i^2=0,$ where $p_i$ are external momenta. In general, infrared
region is not dangerous off shell in component formulation in
Wess--Zumino gauge when we regularize the theory by higher
derivatives.

In such way we come to our main conclusion in this paper. Namely,
${\cal N} =4$ supersymmetric theory has scale-independent
effective action in terms of the dressed effective fields. All the
dependence  on  the dimensionful parameter of ultraviolet
regularization remains in the dressing functions only. This is in
correspondence with direct calculation of anomalous dimensions and
beta function in   ${\cal N} =4$ theory \cite{Avdeev:1980bh}. At
one-loop level, kernels of this scale-independent theory are in
general dilogarithms in momentum space. These dilogarithms are
Fourier transforms of the kernels in position space as given
below.  For example, the correlator of the dressed effective
fields $\tilde{L}$, $\tilde{c}$, and $\tilde{c}$ at one loop level
in any $SU(N)$ gauge theory has, among others, the following
contribution:
\begin{gather}
\langle \tilde{L}^{a}(x)\tilde{c}^{b}(y)\tilde{c}^{c}(z)\rangle
\sim g^2N \frac{1}{((z-y)^2)^2(x-y)^2(z-x)^2}f^{abc},
\label{structure}
\end{gather}
where the dressed effective fields are made of undressed effective
fields convoluted to the dressing functions. The latter are
unspecified but they are parts of the two point proper Green
functions. The terms of the type (\ref{structure}) can be
obtained, for example, by calculating the one-loop $Lcc$ Green
function in the Landau gauge (where $G_c(x) = \del^{(4)}(x)$) and
then using repeatedly the identity
\begin{gather*}
\frac{1}{(2\pi)^4}\int d^4k\, \frac{e^{-ikx}}{k^2 + i\e} =
\lim_{\eta \rar +0} \frac{i}{4\pi^2} \frac{1}{\left[(|x^0|
-i\eta)^2 - {\boldsymbol x}^2\right]}.
\end{gather*}
By the ST identity the correlators (\ref{structure}) are related
to the vertex $KAc$ which in its turn is related to the tree gluon
vertex $AAA.$ The relation of these vertices is dictated by the ST
identity and can be explicitly verified. Thus, the contribution
similar to (\ref{structure}) can be found in the proper correlator
of the three dressed gluons at one-loop level $\langle
\tilde{A}_{\mu}^{a}(x)\tilde{A}_{\nu}^{b}(y)\tilde{A}_{\lambda}^{c}(z)\rangle.$
At the same time, in ${\cal N} =4$ supersymmetry we do not need to
make additional renormalization in two point gluon function to
absorb the rest of infinities from it into renormalization of the
gauge coupling, since the $\b$ function is zero.

We found that the dressed effective fields are the actual
variables of the effective action. The effective action is to be
written in terms of these dressed effective fields. In general, in
non-supersymmetric gauge theory like QCD the dependence on UV
regularization scale will be present inside the correlators of the
dressed effective fields because it is necessary to remove the
dependence on this scale by renormalization of the gauge coupling
constant. In ${\cal N} =4$ supersymmetric theory such a
renormalization does not take place. Thus, the kernels for the
dressed fields do not depend on  scale.  This might make possible
an analysis of these kernels by tools of conformal field theory in
all orders of perturbation theory.

We have shown in this paper that such a scale independent
structure of correlators is a~direct consequence of the
Slavnov--Taylor identity and it is encoded in the  $Lcc$
correlator of the dressed effective fields. In general, by solving
step-by-step the ST identity it is possible to reproduce structure
of all $n$-gluonic proper correlators in terms of dressed
effective fields. Here the question how to define the concept of
scattering may arise. The knowledge of correlators is not
sufficient to define a scattering matrix. Indeed, it follows by
the above construction that these correlators of dressed mean
fields in ${\cal N} =4$ supersymmetry do not have dependence on
any mass parameter, or, stated otherwise, the theory in terms of
dressed mean fields is conformally invariant. It is known that
$S$-matrix in conformal field theory cannot be constructed, since
we do not have any dimensional parameter like mass or scale to
define scattering concepts like typical scattering length, or size
of a meson and so on. The only observables in this theory are
correlators of the gauge invariant operators and their anomalous
dimensions. However, on shell, when we go to the amplitudes, the
scale appears due to infrared on shell divergences, so that
scattering concepts can be introduced in a traditional way
\cite{Bern:2005iz}.

To conclude, in ${\cal  N} =4$ supersymmetric Yang--Mills theory
there are  two types  of scale-independent  correlators. They
are, first, correlators of BPS  gauge invariant operators which
can  be found from AdS/CFT correspondence  and, second,
correlators of the  dressed mean fields,  for example, dressed
gluons in Landau gauge,  that have been  considered in  this
paper.  Correlators of  the  gauge invariant operators  are
surely   gauge  independent.   Non-BPS  operators  are
scale-dependent  and their  anomalous  dimensions are  related to
the infrared   singularities  of  gluon   amplitudes  that   are
on-shell correlators   in  the  momentum   space
\cite{Bern:2005iz}.    As  to correlators of the dressed mean
fields, their gauge dependence is an open question. We make a
conjecture  that all gauge dependence is contained in the
dressing functions  of the dressed  mean fields, and  that the
kernels of the dressed mean fields are gauge invariant.

\subsection*{Acknowledgements}

The work of I.K.\ was supported by Ministry of Education (Chile)
under grant Mecesup FSM9901 and by DGIP UTFSM, by Fondecyt (Chile)
grant \#1040368, and by Departamento de Investigaci\'on de
Universidadad del Bio-Bio, Chillan. The work of G.C.\ and I.S.\
was supported by Fondecyt (Chile) grants \#1010094 and \#1030355,
respectively.  I.K.\ is grateful to Boris Kopeliovich for the
discussions of infrared divergences.

 \LastPageEnding
\end{document}